\documentclass[aps,prb,twocolumn,groupedaddress,showpacs]{revtex4}

\usepackage{graphics}
\usepackage{amssymb}

\begin{document}

\title{Stability of Ca-montmorillonite hydrates:
A computer simulation study}

\author{G. Odriozola}
\email[]{godriozo@imp.mx}

\author{J. F. Aguilar}
\email[]{aguilarf@imp.mx}

\affiliation{Programa de Ingenier\'{\i}a Molecular, Instituto
Mexicano del Petr\'{o}leo, L\'{a}zaro C\'{a}rdenas 152, 07730
M\'{e}xico, Distrito Federal, M\'{e}xico}

\date{\today}
\begin{abstract}
Classic simulations are used to study interlayer structure,
swelling curves, and stability of Ca-montmorillonite hydrates. For
this purpose, $NP_{zz}T$ and $\mu P_{zz}T$ ensembles are sampled
for ground level and given burial conditions. For ground level
conditions, a double layer hydrate having 15.0 $\hbox{\AA}$ of
basal spacing is the predominant state for relative vapor
pressures ($p/p_0$) ranging in 0.6-1.0. A triple hydrate counting
on 17.9 $\hbox{\AA}$ of interlaminar distance was also found
stable for $p/p_0$$=$1.0. For low vapor pressures, the system may
produce a less hydrated but still double layer state with 13.5
$\hbox{\AA}$ or even a single layer hydrate with 12.2 $\hbox{\AA}$
of interlaminar distance. This depends on the established initial
conditions. On the other hand, the effect of burial conditions is
two sided. It was found that it enhances dehydration for all vapor
pressures except for saturation, where swelling is promoted.
\end{abstract}

\maketitle

\section{Introduction}

\label{intro} Clay minerals are negatively charged layer type
aluminosilicates kept together by cations. Since they constitute a
great portion of soils and sedimentary rocks, their study impacts
on podology, geology, geochemistry and ecology. Clays such as
smectite group \cite{Norrish54} have the ability of absorbing
water among clay sheets (interlayer spaces), in some cases
producing a remarkable expansion of the mineral. This expansion is
firstly crystalline (few water layers), and an osmotic regime is
archived for higher water interlaminar contents \cite{Olphen77}.
The study of ions interacting with hydrated clay minerals is of
particular interest, as they rule the swelling capacity of a given
clay.

In engineered settings, clay mineral swelling is a critical factor
in problems such as borehole stability in petroleum extraction,
and in the liner and buffer stability in the containment of
hazardous waste in geoenvironmental technology. One way to
stabilize the shale is by reducing the swelling capacity of the
clay by means of replacing sodium ions with divalent calcium ones
through cation exchange methods
\cite{Marshall49,Norrish54,McEwan80,Olphen87}. Literature reports
indicate that clays containing exchangeable calcium ions swell not
as much as sodium clays, which are known to swell up to form water
clay dispersions\cite{Norrish54,Chatterji79,Olphen87,Sridharan96}.
This makes calcium chloride solutions the preferred internal
phase of most oil-based drilling fluids. Additionally, calcium ions
control water activity. This determines the osmotic movement of
water between the drilling fluid and the formation, which may
reduce the clay swelling when it is correctly handled. Thus, a
detailed knowledge of the stability of the Ca-clay hydrates under
different water activities is a key for an appropriate mud design.

As well as other Ca-montmorillonite simulation studies
\cite{dePablo01b,Greathouse02}, this work deals with the
microscopic mechanisms underlaying Ca-montmorillonite swelling,
but focussing on the stability of the different hydrates in
contact with several reservoirs. These reservoirs differ in
temperature, pressure, and composition (water activity). In order
to fix these variables, the $\mu P_{zz}T$ ensemble is sampled,
where $\mu$ refers to the water chemical potential of the
reservoir. Since it is related to the relative partial water vapor
pressure, plots of interlaminar distances and number of water
molecules against it are constructed. For ground level conditions
(P$=$ 1 atm and T$=298$K), these plots are directly accessed by
experiments, making the comparison easy. Since good agreement was
found, we expect the model to predict the behavior of the system
for other non easily implementable experimental conditions. That
is, for a given average burial depth. Hence, we studied the
behavior of the model under P$=600$ atm and T$=394$K, {\it i.~e.},
for 4 km of depth assuming average gradients of 30 K/km and 150
atm/km. We expect these data to be useful for developing drilling
strategies.

This article is organized as follows. In Sec.~\ref{methods}, we
briefly describe the models and the methodology employed for
performing the simulations. The results are shown in
Sec.~\ref{results}. Finally, Sec.~\ref{summary} summarizes the
main results and extracts some conclusions.

\section{Methodology}
\label{methods}
\subsection{The model}

A montmorillonite clay simulation supercell was constructed by a
$4 \times 2$ replication of the cell given by Skipper {\em et~al.}
\cite{Skipper95a}. The only difference is that in our model the
octahedral oxygen sites have a charge of -1.52$e$, and their
corresponding hydrogen sites have a charge of 0.52$e$. This is for
a better accordance with the TIP4P model \cite{Jorgensen79}, which
was employed for modelling water\cite{Boek95a}. A Wyoming type
montmorillonite was obtained by isomorphous substitutions of
trivalent Al atoms of the octahedral sites by divalent Mg atoms,
and tetravalent Si of the tetrahedral sites by trivalent Al atoms.
The resulting layer counts on dimensions of $21.12 \times 18.28$
$\hbox{\AA}$ in the ($x,y$) plane and a thickness of $6.56$
$\hbox{\AA}$. Water molecules were randomly distributed in the
interlaminar spaces. The negative charge of the clay framework was
balanced by three calcium ions placed in the interlayer midplanes.
The resulting unit cell formula is given by
Ca$_{0.375}$$n$H$_2$O(Si$_{7.75}$Al$_{0.25}$)(Al$_{3.5}$Mg$_{0.5}$)O$_{20}$(OH)$_4$
\cite{dePablo01b}.

\begin{table}
\caption{\label{parameters} Lennard-Jones parameters for
water-clay interactions. }
\begin{ruledtabular}
\begin{tabular}{ccc}
$\;\;$ Sites $\;\;$ & $\epsilon$ (kcal/mol) & $\;\;$ $\sigma$ ($\hbox{\AA}$) $\;\;$ \\
\hline O & 0.1550 & 3.1536 \\
H & 0.0000 & 0.0000 \\
Si & 3.150 & 1.840 \\
Al & 3.150 & 1.840 \\
Mg & 3.150 & 1.840 \\
\hline
\end{tabular}
\end{ruledtabular}
\end{table}

The water-clay interactions are taken from Boek \cite{Boek95b}
{\em et~al.} Here, the total interaction potential is given by a
Coulombian plus a Lennard-Jones contribution,
\begin{equation}\label{pot}
U_{ij}\!=\!\sum_{a,b}\! \left[ \frac{q_aq_b}{r_{ab}}+ 4
\epsilon_{ab} \left[ \left( \frac{ \sigma_{ab}}{r_{ab}}
\right)^{12} - \left( \frac{ \sigma_{ab}}{r_{ab}} \right)^{6}
\right] \right]
\end{equation}
where subindexes $i$ and $j$ are for molecules, and $a$ and $b$
run over all sites of each molecule. $q_{a}$ and $q_{b}$ are the
corresponding site charges, $\epsilon_{ab}$ and $\sigma_{ab}$ are
site-to-site specific Lennard-Jones parameters, and $r_{ab}$ is
the inter-site distance. The site-to-site Lennard-Jones parameters
are given by the Lorentz-Berthelot rules
\begin{equation}
\sigma_{ab}=\frac{\sigma_a + \sigma_b}{2},
\end{equation}
\begin{equation}
\epsilon_{ab}=\sqrt{\epsilon_a \epsilon_b}
\end{equation}

The corresponding Lennard-Jones parameters for different species
are given in Table \ref{parameters}. Parameters for Si were taken
from Marry {\em et~al.} \cite{Levesque02}, and parameters for Al
and Mg were assumed to be equal to those of Si.

The Ca-O and Ca-H interactions are based on the ones proposed by
Bounds \cite{Bounds85}, since they produce Ca-TIP4P radial
distribution functions in agreement with available experimental
data and close to hybrid quantum mechanics/ molecular mechanics
(QM/MM) results. That is, the Ca-O radial distribution function
peaks at 2.54 $\hbox{\AA}$ leading to a first shell oxygen
coordination number of 9.3, while the experimental results are
close to 2,46 $\hbox{\AA}$ \cite{Jalilehvand01} and a wide range
of coordination numbers turn into 6.0-10.0
\cite{Jalilehvand01,Hewish82,Probst85,Licheri76}. On the other
hand, hybrid QM/MM simulations performed at DFT level (LANL2DZ
basis sets) lead to 2.51 $\hbox{\AA}$ of Ca-O distance and a
coordination number of 8.1 \cite{Schwenk01}. The pair potential
reads
\begin{eqnarray}
U_{Ca-H_2O} \!\!&=& \!\! A_{CaO}\exp{(-b_{CaO}r_{CaO})}-C_{CaO}/r_{CaO}^4 \nonumber \\
& & \!\! -D_{CaO}/r_{CaO}^6+A_{CaH}\exp{(-b_{CaH}r_{CaH_1})} \nonumber \\
& & \!\! +A_{CaH}\exp{(-b_{CaH}r_{CaH_2})} \label{Bounds},
\end{eqnarray}
with $A_{CaO}=$ 37146.0 kcal mol$^{-1}$, $b_{CaO}=$ 2.9902
$\hbox{\AA}^{-1}$, $C_{CaO}=$ 1578.6 kcal $\hbox{\AA}^4$
mol$^{-1}$, $D_{CaO}=$-2185.3 kcal $\hbox{\AA}^6$ mol$^{-1}$,
$A_{CaH}=$ 8212.0 kcal mol$^{-1}$ and $b_{CaH}=$ 3.7234
$\hbox{\AA}^{-1}$.

Since it is crucial for the hybrid Monte Carlo (HMC) simulations
to keep the energy fluctuations as low as possible in order to
enlarge the acceptation rate \cite{Mehlig92}, it is convenient to
avoid employing relatively long range pair potential
contributions, such as $\sim r^{-4}$, if no Ewald treatment is
applied on them. Thus, we refitted to equation \ref{Bounds} the
following expression
\begin{eqnarray} \label{fit}
U_{Ca-H_2O} \!\!&=& \!\! A_{CaO}\exp{(-b_{CaO}r_{CaO})}-C_{CaO}/r_{CaO}^6 \nonumber \\
& & \!\! +A_{CaH}\exp{(-b_{CaH}r_{CaH_1})} \nonumber \\ & & \!\!
+A_{CaH}\exp{(-b_{CaH}r_{CaH_2})},
\end{eqnarray}
by employing a Levenberg-Marquardt algorithm, considering several
Ca-H$_{2}$O configurations. The procedure yields $A_{CaO}=$
229184.5 kcal mol$^{-1}$, $b_{CaO}=$ 3.3626 $\hbox{\AA}^{-1}$,
$C_{CaO}=$ 15616.5 kcal $\hbox{\AA}^6$ mol$^{-1}$, $A_{CaH}=$
2417.0 kcal mol$^{-1}$ and $b_{CaH}=$ 3.0382 $\hbox{\AA}^{-1}$. We
observed that by avoiding the $\sim r^{-4}$ term, the acceptance
rate enlarges more than three times for a small (inner) time step
of 8$\times$10$^{-4}$ ps, this being the typical value. To
check the reliability of this Bound based Ca-water potential,
$NPT$ simulations with 216 water molecules, a calcium cation, and
two chloride anions were performed. From the radial distribution
functions we observed that the first and second oxygen shells are
situated at 2.53 and 4.43 $\hbox{\AA}$ respectively, while for
Ca-H they are at 3.05 and 5.05 $\hbox{\AA}$. The coordination
number for the first oxygen coordination shell is 8.75. These
values are in agreement with experimental data and are casually
closer than Bound's potential results to those obtained
by Schwenk {\em et~al.} \cite{Schwenk01} Additionally, results
also compare well with those reported by D.~G.~Bounds
\cite{Bounds85}. Therefore, expression \ref{fit} seems suitable
for describing the Ca-TIP4P interactions.

Periodic boundary conditions were imposed on the three space
directions. The electrostatic interactions $\sim r^{-1}$ are
computed using the Ewald summation method, and a spherical cutoff
equal to half the smallest box side is set for the short range
interactions. Standard corrections for the short range
interactions were considered \cite{Frenkel}.

\subsection{Simulations}
\label{simulation} Our simulation methodology is based on previous
work \cite{Odriozola_jcp} thus it is not discussed in detail here.
The simulations were performed employing a HMC method
\cite{Mehlig92,Odriozola_jcp}. A reversible multiple time scale
algorithm \cite{Tuckerman92} is employed as discretization scheme.
The long time step is set up to 8 times the short time step, and
the short time step is chosen to obtain an average acceptance
probability of 0.7 \cite{Mehlig92}. To keep time correlations as
low as possible, a new configuration is generated each 10
integration steps. The probability to accept this new
configuration is
\begin{equation}
P=\min\{1,\exp(-\beta \Delta \mathcal{H})\}
\end{equation}
where $\Delta \mathcal{H}$ is the difference between the new and
previous configuration Hamiltonians, and $\beta$ is the inverse of
the thermal energy.

For sampling in the $NP_{zz}T$ ensemble the stress normal to the
surface of the clays, $P_{zz}$, is kept constant. Thus, only
volume fluctuations in the $z$-direction are allowed, and the
probability for accepting a new box configuration is given by
\begin{equation}
P\!=\! \min\{ 1,\exp [ - \beta ( \Delta \mathcal{U}+P_{zz} \Delta
V \! - N \beta ^{-1} \! \ln (V_n/V_o)) ] \}
\end{equation}
Here, $\Delta \mathcal{U}$ is the change in the potential energy,
$\Delta V$ is the volume change, $N$ is the total number of
molecules, and $V_n$ and $V_o$ are the new and old box volumes,
respectively \cite{dePablo01}.

\begin{figure}
\resizebox{0.4\textwidth}{!}{\includegraphics{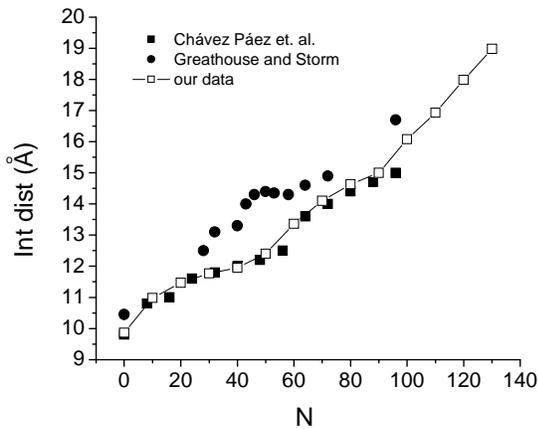}}
\caption{\label{NPT_swell_sup} Interlaminar distance as a
function of the number of water molecules per interlaminar space
for ground level conditions.}
\end{figure}

The $\mu P_{zz}T$ ensemble is sampled by simply performing
particle movements, insertions and deletions, and box changes as
in typical $NVT$, $\mu VT$, and $NP_{zz}T$ samplings
\cite{Odriozola_jcp}. Water insertions and deletions were
performed by Rosenbluth sampling \cite{Hensen01,Odriozola_jcp}.
Different types of movements are called just like explained
elsewhere \cite{Odriozola_jcp}. This algorithm provides a good way
of sampling this kind of systems, since only one run is necessary
to obtain the equilibrium state at reservoir conditions. This
contrasts with the more frequently used $\mu VT$ sampling, that
leads to similar information after a large number of simulation
runs \cite{dePablo01b,dePablo04,dePablo01,Hensen01,Hensen02}. It
should be also mentioned that in typical x-ray experiments the
ambient vapor pressure, the total pressure, and the temperature
are controlled, but not the water content of the interlaminar
space or the interlaminar distance. These are exactly the same
variables that are setup in a $\mu P_{zz}T$ simulation, which
consequently allows for a fair comparison. On the other hand,
$NP_{zz}T$ and $\mu VT$ ensembles may force the system to pass
trough equilibrium states that are not produced in real
experiments. This is the case of water contents equivalent to 55
water molecules per interlaminar space or interlaminar distances
of 13 $\hbox{\AA}$ for sodium montmorillonite, since they do not
correspond to a single or to a double water hydrate. Nevertheless,
one may artificially produce these states by simply fixing $N$ or
$V$ in $NP_{zz}T$ or $\mu VT$ sampling. Measuring the water
chemical potential for the first case \cite{Odriozola_lan}, or the
pressure for the second \cite{dePablo01}, large values of chemical
potential and pressure are obtained. In fact, these variables show
oscillations when obtained as a function of $N$ or $V$, indicating
that several equilibrium states are possible
\cite{Odriozola_lan,dePablo01}. These measurements also suggest
the existence of energetic barriers that separates the single,
double, and triple layer hydrates. Since in $\mu P_{zz}T$ sampling
the interlaminar distance and the number of water molecules are
free variables, these states are simply avoided, splitting the
phase space in two (or more) regions. For these cases, two (or
more) different equilibrium states appear, which are accessed by
handling initial conditions. This way, hysteresis cycles are
naturally obtained.

\section{Results}
\label{results}

\begin{figure}
\resizebox{0.45\textwidth}{!}{\includegraphics{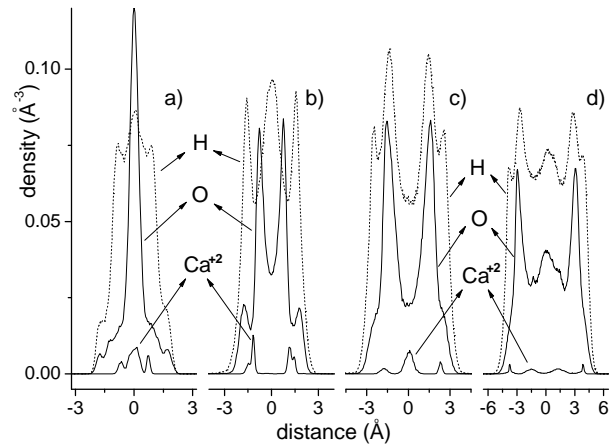}}
\caption{\label{perfil_40-60-90-120} Oxygen, hydrogen and
calcium density profiles of the interlaminar space for ground
conditions. The water amount was fixed to 40, 60, 90 and 120
molecules per interlaminar space, from left to right,
respectively.}
\end{figure}

The swelling curve obtained for ground level conditions, {\it
i.~e.}~P=1 atm and T=298 K, is shown in figure
\ref{NPT_swell_sup}. Here, each data point is obtained by
performing a $NP_{zz}T$ simulation and hence, by fixing the number
of water molecules. This figure also includes the data reported by
Ch\'{a}vez-P\'{a}ez {\em et~al.} \cite{dePablo01} and by
Greathouse and Storm \cite{Greathouse02}. As can be seen, our data
compare very well with those of Ch\'{a}vez-P\'{a}ez {\em et~al.}
Since methodologies and models are similar this is something
expected. Larger differences are seen between our data and those
reported by Greathouse and Storm, who employed the Lennard-Jones
type Ca-O potential given by Aquist \cite{Aquist90}. These
differences point out to a general weakness associated with the
use of classical force fields, which may be overcome by employing
ab initio molecular dynamics simulations. This was successfully
done by Boek {\em et~al.} \cite{Boek03}

Systems counting on 40, 60, 90 and 120 water molecules (wm) per
interlaminar space produce 12.0, 13.4, 15.0, and 17.9 $\hbox{\AA}$
of interlaminar distance. The first one corresponds to a single
water layer formation. The last two are frequently observed in
experiments and should correspond to double and triple water layer
hydrates, while the interlaminar distance of 13.4 $\hbox{\AA}$ is
sometimes obtained experimentally for small vapor pressures
\cite{Mooney52,Hendricks40,Sato92}. The corresponding oxygen,
hydrogen, and calcium profiles are shown in figure
\ref{perfil_40-60-90-120}. Oxygen peaks of this figure make clear
that the structures correspond to a single water layer hydrate, to
two double layer hydrates, and to a three layer hydrate, from left
to right. It also shows that the double layers structures differ
on their hydrogen and calcium profiles. These calcium profiles pass from two
double peaks located close to the clay layers, suggesting the
formation of different types of inner-sphere complexes, to a
structure having a very important middle peak, indicating the
formation of outer-sphere complexes. This tendency for calcium
ions to detach from the surface with increasing water content agree
with the predictions of Greathouse and Storm \cite{Greathouse02}.

\begin{figure}
\resizebox{0.44\textwidth}{!}{\includegraphics{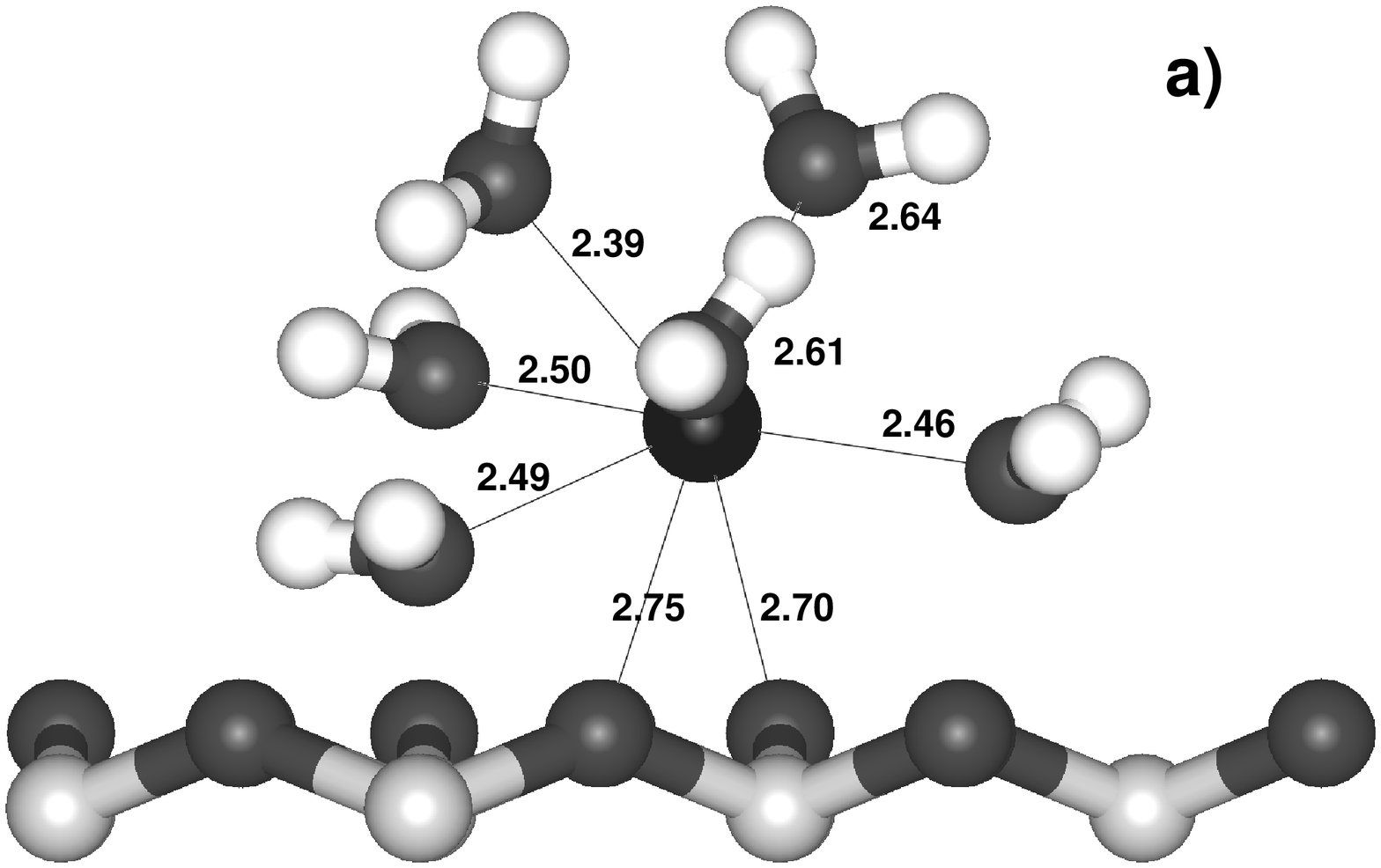}}
\resizebox{0.42\textwidth}{!}{\includegraphics{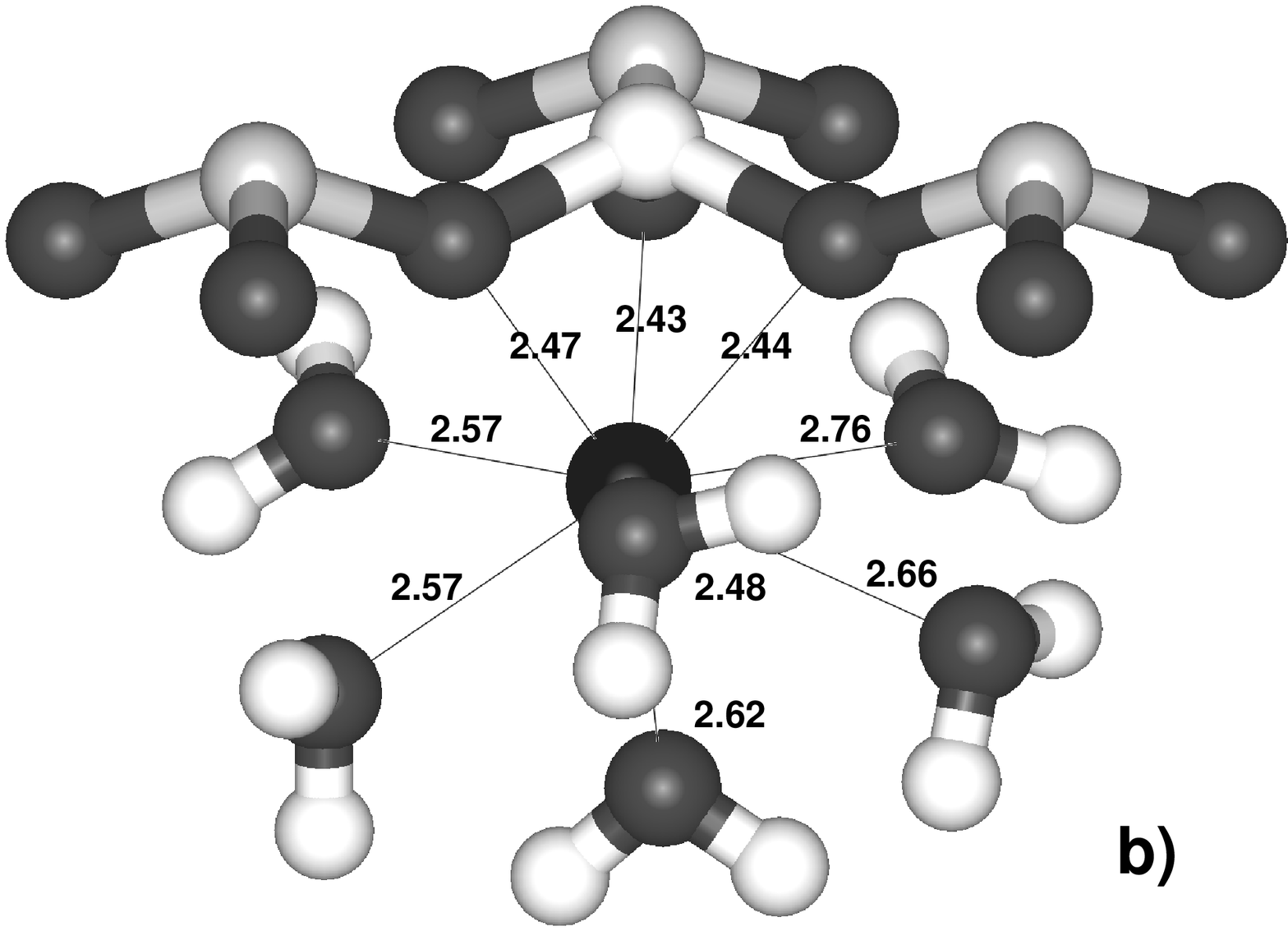}}
\resizebox{0.39\textwidth}{!}{\includegraphics{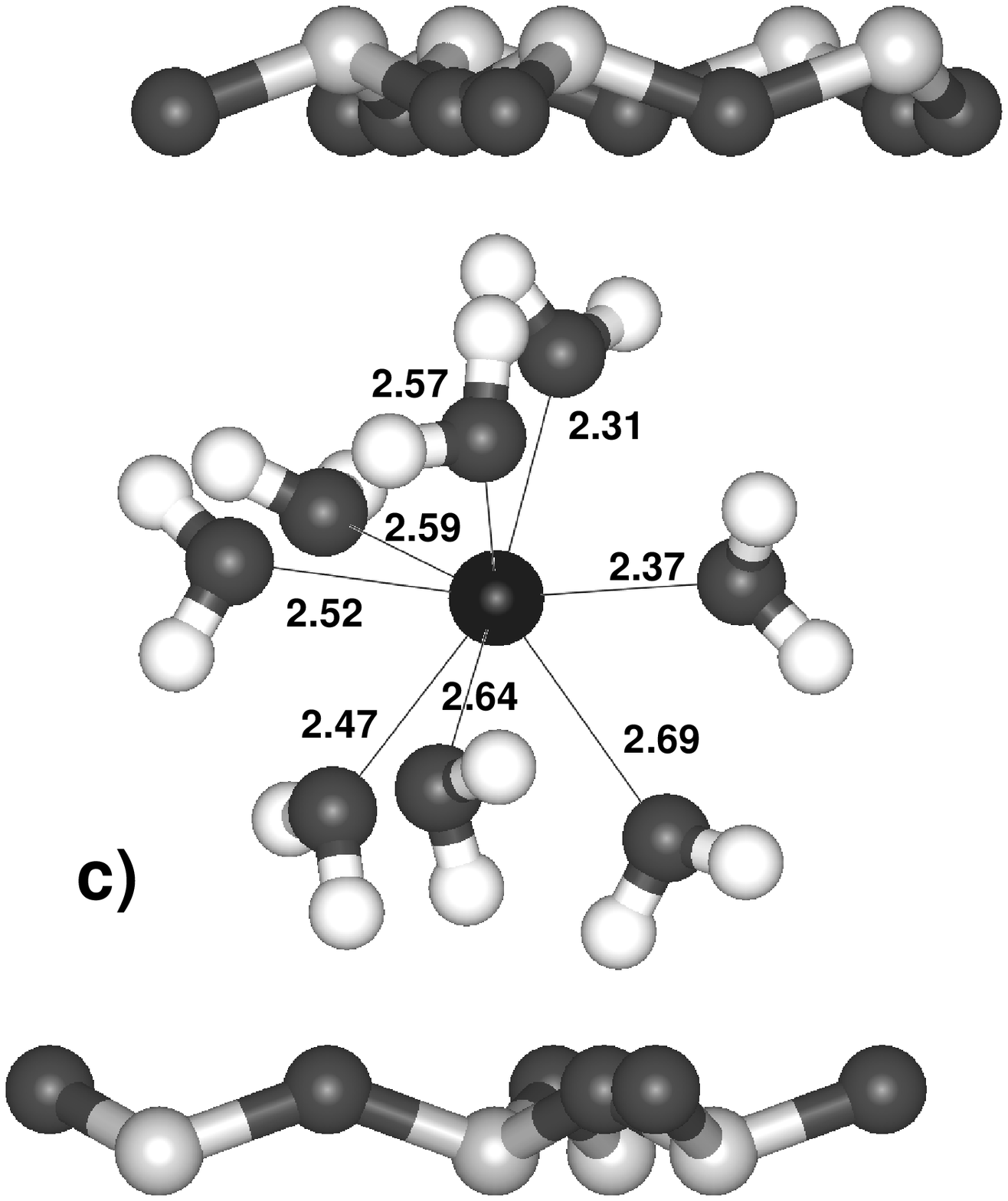}}
\caption{\label{complexes} Zoom in of calcium ions (black)
obtained from equilibrium configurations. Only water molecules
counting on Ca-O distances smaller than 3.0 $\hbox{\AA}$ are
shown. H are represented by white, O by dark-gray, and Si atoms by
light-gray spheres. Distances in the figure are given in
$\hbox{\AA}$. a) Ca$^{++}$ forming an inner-sphere surface
complex. b) Ca$^{++}$ forming an inner-sphere surface complex that
involves a tetrahedral substitution of a silicon by an aluminium
(white), and c) Ca$^{++}$ forming an outer-sphere complex.}
\end{figure}

To confirm the presence of inner and outer-sphere complexes some
snapshots were analyzed. From them figure \ref{complexes} was
built, where only those water molecules having Ca-O distances
smaller than 3.0 $\hbox{\AA}$ are shown. Two inner-sphere
complexes are seen, \ref{complexes} a) and b), and an outer-sphere
complex, \ref{complexes} c). The complex shown in figure
\ref{complexes} a) has no tetrahedral substitution involved,
whereas in figure \ref{complexes} b) the calcium ion is attached
to a tetrahedral substitution. As can be seen, two oxygens
coordinate with the ion for the first case, and the three
surrounding the aluminium atom in the second. In both cases, 6 wm
complete the inner sphere shell. In addition, distances from clay
oxygens are much smaller when the tetrahedral substitution is
involved. These different coordination distances explain the
calcium double peaks seen in figure \ref{perfil_40-60-90-120} b).
On the other hand, 8 wm form the inner sphere shell in case of
outer-sphere complexes, in agreement with
simulations\cite{Greathouse02} and experimental evidence
\cite{Slade85}. For double layer hydrates, these complexes are
situated close to the interlayer midplane, leading to Ca-O
distances for clay oxygens in the range of 4.3-5.0 $\hbox{\AA}$
(these distances are not highlighted in the figure). It should be
noted that they are similar to the Ca-O distances found for the
water second shell that surrounds calcium in bulk water. Hence,
this may enhance the double layer stability.

In order to build swelling curves as a function of the water vapor
pressure of a reservoir in contact with the system, the $\mu
P_{zz}T$ ensemble was sampled. The relationship between the
chemical potential and the vapor pressure is $\beta \mu$ $=$
$\beta \mu_0$ $+ \ln(p/p_0)$, where $p_0$ is the vapor pressure at
equilibrium with liquid water whose chemical potential is $\mu_0$,
and $p$ is the vapor pressure. For the TIP4P model, 1 atm, and 298
K, we employed $\beta \mu_0$$=$ -17.4
\cite{dePablo01,Odriozola_jcp}. In case of 600 atm and 394 K the
employed value is -13.4 \cite{Odriozola_jcp}.

\begin{figure}
\resizebox{0.45\textwidth}{!}{\includegraphics{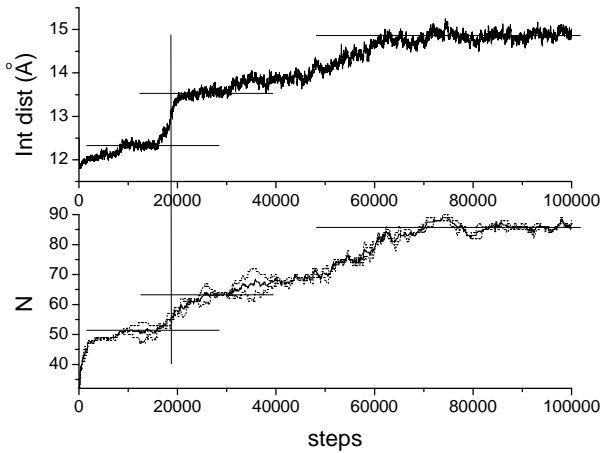}}
\caption{\label{run_p08} Evolution of the interlaminar
distance and water content with the number of simulation steps.
For the lower plot, dotted lines are the water content for each
interlaminar space and the solid line is the average. Initial
conditions are close to the dehydrated state, p$=$0.8$p_0$, and
ground level conditions were imposed. Horizontal and vertical
lines are just a guide to the eye.}
\end{figure}

A sampling example from the $\mu P_{zz}T$ ensemble is shown in
figure \ref{run_p08}. Here, the interlaminar distance, the water
amount of each interlaminar space, and the average amount of water
are plotted against the simulation step. For this particular case
the simulation was started from an almost dehydrated state in
contact with a reservoir having $p$$=$$0.8p_0$, and for ground
level conditions. It is seen how water molecules enter the system
thickening the interlaminar distance. It is also noticed that once
the system yields some of the interlaminar structures shown in
figure \ref{perfil_40-60-90-120}, it shows certain resistance to
swell to another state. For example, the resistance at 12.4
$\hbox{\AA}$, corresponding to a single water layer hydrate, is
overcome once the simulation has spent close to 8000 steps.
Immediately after overcoming this resistance, the interlaminar
distance jumps from 12.4 to 13.5 $\hbox{\AA}$. This means that the
interlaminar space rearranges from the single water layer shown in
figure \ref{perfil_40-60-90-120} a) to the double water layer
corresponding to figure \ref{perfil_40-60-90-120} b). Clearly,
this rearrangement implies overcoming a sort of collective
potential barrier. Something similar happens with the transition
from the double layer hydrate at 13.5 $\hbox{\AA}$ towards the
structure shown in figure \ref{perfil_40-60-90-120} c), although
the rearrangement seems to be gradual. Here, most ions leave the
surface to locate close to the interlaminar midplane. Finally, the
system yields a stable state close to 15.0 $\hbox{\AA}$ (stable at
least for the finite number of considered steps, but confirmed by
other runs having starting configurations closer to this final
state).

We consider that the resistances found for rearranging the
interlaminar structures point to local free energy minima, and
their relative depths under the given conditions should be
directly related to the number of steps the simulation spends on
them. However, due to the stochastic character of this number,
several runs should be performed to obtain some reliable averages.

\begin{figure}
\resizebox{0.45\textwidth}{!}{\includegraphics{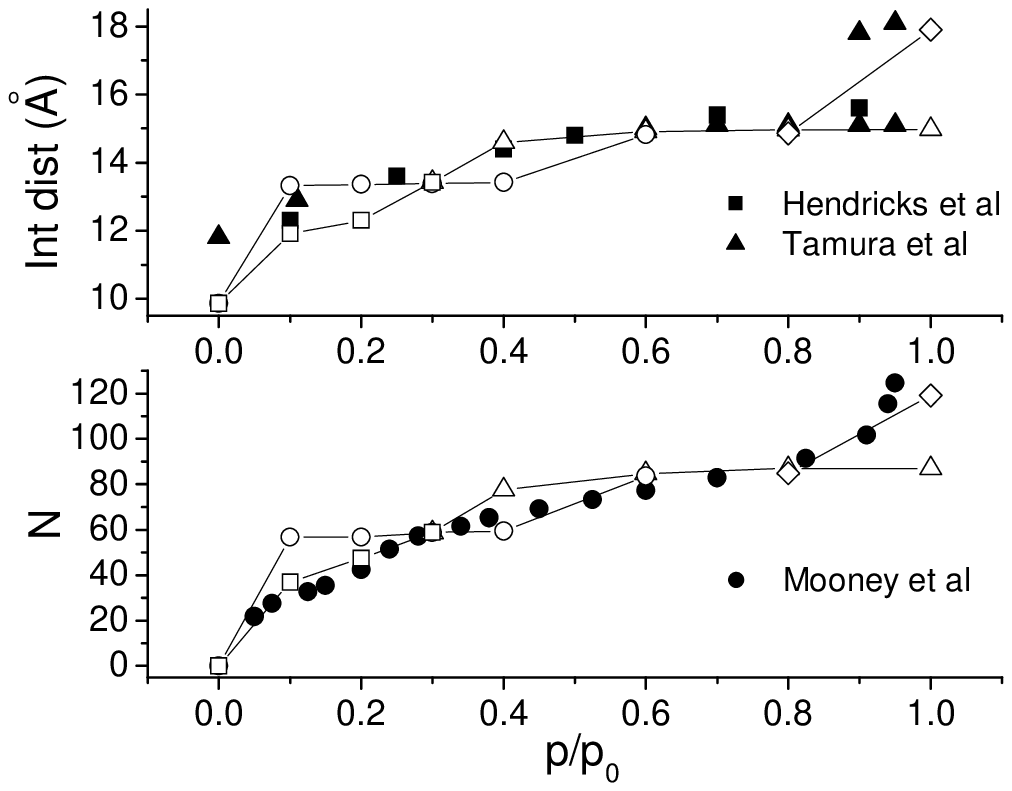}}
\caption{\label{GC_swell_sup} Interlaminar distance and number
of water molecules per interlaminar space as a function of the
vapor pressure for ground level conditions. Symbols {\tiny
$\Box$}, $\circ$, {\tiny $\triangle$}, and {\tiny $\lozenge$}
correspond to initial conditions of 10 wm - 12.0 $\hbox{\AA}$, 60
wm - 15.0 $\hbox{\AA}$, 90 wm - 16.0 $\hbox{\AA}$, and 120 wm and
18.5 $\hbox{\AA}$ of interlaminar distance, respectively.}
\end{figure}

The swelling curves for ground level conditions are shown in
figure \ref{GC_swell_sup}. To generate them, four starting
configurations were considered. These are: an almost dehydrated
state having 10 wm and 12.0 $\hbox{\AA}$ of interlaminar space, 60
wm - 15.0 $\hbox{\AA}$, 90 wm - 16.0 $\hbox{\AA}$, and 120 wm -
18.5 $\hbox{\AA}$ of interlaminar space. These configurations
produce single, double, and triple layer hydrates in a few
simulation steps. If the fixed conditions are consistent with
these states, both, water content and interlaminar distance just
fluctuate around certain mean values. On the contrary, if vapor
pressure is not consistent with the hydrate, this is destabilized,
producing another number of water layers. This is the case shown
in figure \ref{run_p08}. Limiting cases are the dehydrated state
(no water molecules remain in the system) and the fully hydrated
state (where the simulation leads to an ever increasing number of
water molecules).

As expected, independently of the initial conditions, the
simulations for $p=0$ yield the dehydrated state, counting on an
interlaminar distance of 9.9 $\hbox{\AA}$, in agreement with the
$NP_{zz}T$ results. On the other hand, and also no mattering what
the established initial conditions are, for $p$ $\geq$ $0.6p_0$ a
double layer hydrate is always obtained, except for a saturated
vapor pressure where a triple layer hydrate is also possible. The
structure of this double layer hydrate is consistent with that
shown in figure \ref{perfil_40-60-90-120} c). It has a water
content close to 87 wm and an interlaminar space about 15.0
$\hbox{\AA}$. On the other hand, the triple layer hydrate
structure looks like that shown in figure
\ref{perfil_40-60-90-120} d) and has 17.9 $\hbox{\AA}$ of
interlaminar space and close to 119 wm.

For vapor pressures ranging in 0.1 - $0.4 p_0$, things are more
complicated. Here, two equilibrium states were observed for
$p$$=$0.4, 0.2, and 0.1$p_0$. These are: two different double
layer hydrates for $p$$=$0.4$p_0$; and a single and a double layer
hydrates for $p$$=$0.2 and 0.1$p_0$. These two double layer
hydrates are those shown in figure \ref{perfil_40-60-90-120} b)
and c). As already mentioned, they differ on their water content
and on their water-ion structure. The single hydrate shows
interlaminar distances close to 12.2 $\hbox{\AA}$. This hydrate was
found stable just in the range of 0.1 - $0.2 p_0$, and the double
layer hydrate with the smallest interlaminar distance is stable
for $p=$ 0.1 - $0.4 p_0$. This completes the description for the
swelling of Ca-montmorillonite hydrates, showing two closed
hysteresis loops for small water vapor pressures.

The obtained data are in good agreement with experimental results.
In order to clearly see it, figure \ref{GC_swell_sup} includes the
experimental data obtained by Hendricks {\em et~al.}
\cite{Hendricks40}, Tamura {\em et~al.} \cite{Tamura00} and Mooney
{\em et~al.} \cite{Mooney52}. Here, not only the interlaminar
distances for the single, double, and triple hydrates are
reasonably matched but also their relative vapor pressure range.
For example, the experimental basal spacing distances for the
single layer hydrate range in 11.9-12.5 $\hbox{\AA}$; 15.0-15.2
$\hbox{\AA}$ are the values reported for the double layer hydrate;
and 17.7-18.1 $\hbox{\AA}$ are those reported for the triple
hydrate \cite{Tamura00,Cuadros97,Mooney52}. Moreover, they even
seem to support the interlaminar distances obtained for the double
layer hydrate shown in figure \ref{perfil_40-60-90-120} b).
Therefore, values that range in 12.8-13.8 $\hbox{\AA}$ for small
water vapor pressures \cite{Mooney52,Hendricks40,Sato92} may
correspond to our double layer of 60 wm per interlaminar space.

The number of water molecules contained in the system also match
the data given by Mooney {\em et~al.} \cite{Mooney52}. We should
mention here that for converting their data from water content to
number of water molecules, it was assumed that 96 wm correspond to
300 mg H$_2$O/g clay \cite{Boek95a,Greathouse02,Newman}. The data
plotted in figure \ref{GC_swell_sup} under the label Mooney {\em
et~al.} were obtained this way. As can be seen, the agreement is
remarkable.

\begin{figure}
\resizebox{0.45\textwidth}{!}{\includegraphics{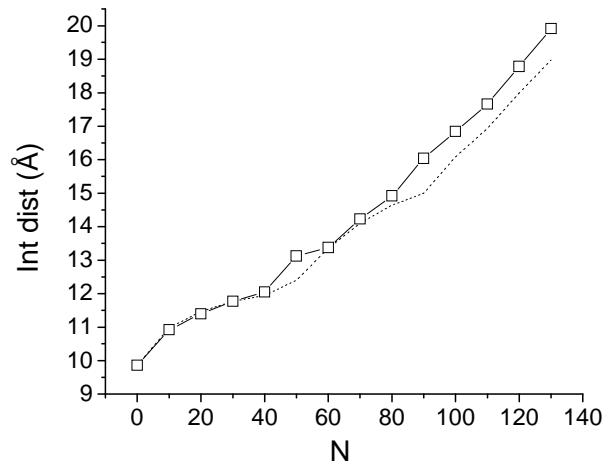}}
\caption{\label{NPT_swell_poz} Interlaminar distance as a
function of the number of water molecules per interlaminar space
for burial conditions. For comparison, the dotted line corresponds
to ground conditions (from figure \ref{NPT_swell_sup}).}
\end{figure}

The swelling curve obtained for burial conditions and by means of
$NP_{zz}T$ sampling is shown in figure \ref{NPT_swell_poz}. We
also include the one obtained for ground conditions, for
comparison. It is seen that for small water contents, both curves
almost coincide. As water content increases, the curve obtained
for burial conditions produces larger interlaminar spaces. Another
effect is that the first plateau shortens, and a double layer
formation is observed for water contents as low as 50 wm. Here,
the first double layer plateau ranges in 50-60 wm, and there seems
to be a second in the range 70-80 wm. This plateau displacement is
due to the large effective volume the water molecules occupy for
burial depth, being just a consequence of the larger temperature
\cite{Odriozola_lan}.

The structures of the systems that correspond to the different
plateaus are similar to those already shown in figure
\ref{perfil_40-60-90-120}. The main difference is that profile
peaks widen and shorten due to higher thermal energy. This was
also observed by means of experiments \cite{Skipper00}. Something
similar happens with the Ca-O radial distribution functions,
although the first shell coordination numbers remain quite the same.

\begin{figure}
\resizebox{0.45\textwidth}{!}{\includegraphics{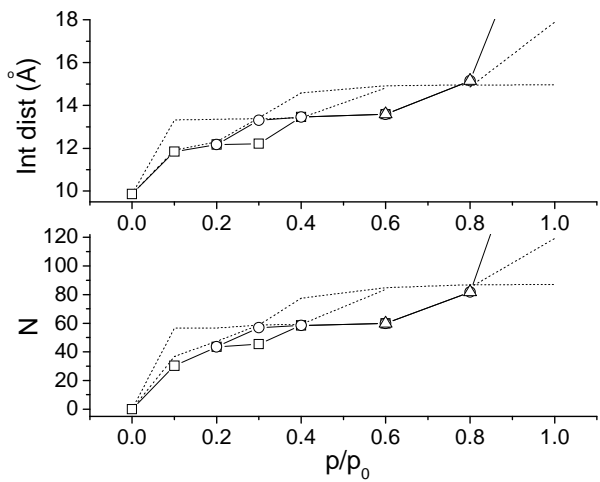}}
\caption{\label{GC_swell_poz} Interlaminar distance and number
of water molecules per interlaminar space as a function of the
vapor pressure for burial conditions. Symbols {\tiny $\Box$},
$\circ$ and {\tiny $\triangle$} correspond to initial conditions
of 10 wm - 12.0 $\hbox{\AA}$, 60 wm - 15.0 $\hbox{\AA}$, and 90 wm
- 16.0 $\hbox{\AA}$ of interlaminar distance, respectively.}
\end{figure}

Figure \ref{GC_swell_poz} shows the swelling curves obtained for
burial conditions. Just like figure \ref{GC_swell_sup}, the water
amount is not fixed since sampling was performed in the $\mu
P_{zz}T$ ensemble. Therefore, in general, several different
equilibrium states appear as a function of the imposed reservoir's
vapor pressure, which are accessed by handling initial conditions.
Figure \ref{GC_swell_poz} also includes the results for ground
level conditions with dotted lines, for an easier comparison. To
begin with, let us focus on the data obtained by starting from an
almost dehydrated state. As expected, the full dehydration is only
obtained for $p$$=$$0$. This has a very similar interlaminar
distance than the one obtained for ground level conditions. For
0.1$p_0$ $\leq$ $p$ $\leq$ 0.3$p_0$, the single water hydrate is
yielded. This contrasts with the vapor pressure range obtained for
the single hydrate at ground level conditions, which is shorter.
This suggests that dehydration is favored for burial conditions.
This hydrate counts on an interlaminar distance similar to that
obtained for ground level conditions, although with a smaller
water content. Again, this points towards the larger effective
volume occupied by water molecules at higher temperatures. It is
also seen that for a vapor pressure of $p$$=$0.4$p_0$, the system
produces a double layer hydrate having 13.3 $\hbox{\AA}$ of
interlaminar distance and 58 wm. This signatures the end of the
single water layer domain.

The results for an initial condition of 60 wm and 15.0
$\hbox{\AA}$ of interlaminar distance are as follows. For $p$ $=$
0.1-0.2$p_0$ the double layer hydrate is destabilized and a single
layer hydrate is produced. In the range 0.3$p_0$ $\leq$ $p$ $\leq$
0.6$p_0$, this initial condition leads to a double layer hydrate
similar to that shown in figure \ref{perfil_40-60-90-120} b). This
state has an interlaminar distance close to 13.3 $\hbox{\AA}$ and
a water content of 58 wm. Our data indicate that for $p$ $=$ 0.4
and 0.6 $p_0$ it turns into the only stable state. Hence, for $p$
$=$ 0.6 $p_0$ the structure \ref{perfil_40-60-90-120} c) obtained
under ground level conditions is also destabilized in favor of the
structure \ref{perfil_40-60-90-120} b). All this points to a
dehydration process occurring at burial depths as well.

The initial condition of 90 wm and 16.0 $\hbox{\AA}$ of
interlaminar distance produces the double layer hydrate having
15.0 $\hbox{\AA}$ of interlaminar distance only for $p$ $=$ 0.8$p_0$. For a
saturated vapor pressure the hydrate becomes unstable and a full
hydration is observed. On the other hand, it dehydrates for $p$
$=$ 0.6$p_0$. Therefore, it seems that burial conditions favor
dehydration for all vapor pressures except for saturation, that,
on the contrary, promotes swelling.

\section{Conclusions}
\label{summary}

Ca-montmorillonite hydrates were studied by means of $NP_{zz}T$
and $\mu P_{zz}T$ simulations. Interlaminar structures and
swelling curves for a fixed amount of water were analyzed by
$NP_{zz}T$ sampling, whereas the $\mu P_{zz}T$ ensemble was used
to build swelling curves as a function of the reservoir relative
vapor pressure. Both ground level and burial conditions were
considered.

Results indicate that under ground level conditions four
interlayer structures are possible. For small relative vapor
pressures, a single or a double layer hydrate with a low water
content are obtained, counting on 12.2 and 13.4 $\hbox{\AA}$ of
interlaminar distance, respectively. For larger relative vapor
pressures, a double hydrate with a higher water content and 15.0
$\hbox{\AA}$ of interlaminar distance is obtained. Finally, a
three layer hydrate with 17.9 $\hbox{\AA}$ of interlaminar
distance was detected for water vapor saturation. All these data
well agree with experimental results.

It was observed that the more hydrated the system becomes, more
ions fully hydrate to form outer-sphere complexes. In other words,
inner-sphere complexes are mostly observed for low water contents.
It should be mentioned that these outer-sphere complexes in double
layer hydrates present part of their second water shell
substituted by oxygen atoms of the two adjacent clay layers. This
aids to counterbalance the expanding pressure the water molecules
exert on the clay sheets. On the other hand, calcium ion lowers
the interlaminar space water activity, favoring the entrance of
water thus producing denser systems. This causes higher expanding
pressures to deal with. Hence, it was not a priory clear if
calcium ions were going to produce stable hydrates under all
environments.

Burial conditions enhance dehydration for all vapor pressures
except for saturation. This last case was found to provoke
swelling. We should mentioned that a real reservoir with a
saturated vapor pressure is not likely to occur, since there are
always dissociated electrolytes that lower water activity. Thus,
one can not expect this extreme case to happen in a real
reservoir. For example, Wang {\em et~al.}\cite{Wang96} were
able to relate the dehydration temperature of montmorillonite
in calcium solutions with its water activity, finding that
relatively small quantities of calcium chloride produce a large
enough drop of the vapor pressure to considerably lower
dehydration temperature. In fact, this explains the extensive use
of calcium chloride as the internal phase of oil-based drilling
fluids. From figure \ref{GC_swell_poz}, it is concluded that a
drop of vapor pressure not only prevents swelling but also favors
dehydration, indeed. Therefore, we consider that the most
important role of calcium ion as swelling inhibitor is just to
lower the water activity of the reservoir in contact with the
montmorillonite hydrate system.

\section{Acknowledgments}
This research was supported by Instituto Mexicano del Petr\'{o}leo
grant D.00072.

%\bibliographystyle{apsrev}
%\bibliography{Clay}

\end{document}